\documentstyle[12pt,epsfig]{article}
\begin{document}
\title{Determination of the scalar glueball mass in QCD sum rules}
\author {{\small Tao Huang$^{1,2}$, HongYing Jin$^2$ and Ailin Zhang$^2$}\\
{\small $^1$ CCAST (World Laboratory), P. O. Box 8730, Beijing, 100080}\\ 
{\small $^2$ Institute of High Energy Physics, P. O. Box 918, Beijing, 100039, P. R.
China}}
\date {}
\maketitle
\begin{center}
\begin{abstract}
The $0^{++}$ glueball mass is analyzed in the QCD sum rules. We show that in
order to determine the $0^{++}$ glueball mass by using the QCD sum rules
method, it is necessary to clarify the following three ingredients: (1) to
choose the appropriate moment with acceptable parameters which satisfy all
of the criteria; (2) to take account into the radiative corrections; (3)
to estimate an additional contribution to the glueball mass from the lowest
lying ${\bar q}q$ resonance. We conclude that it is the key point to choose
suitable moments to determine the $0^{++}$ glueball mass, the radiative
corrections do not affect it sensitively and the composite resonance have a
little effect on it.
\end{abstract}
\end{center}
\section{Introduction}
\indent
\par
The self-interaction among gluons is a distinctive feature in the QCD
theory. It may lead to build bound gluon states, glueballs. Thus discoving
of the glueball will be a direct test to the QCD theory. Although 
there are several glueball candidates experimentally, 
there is no conclusive evidence on them.
People recently pay particular attention to two scalar states: $f_0$(1500)(J=0)
\cite{cbc} and 
$f_J$(1710) (J=0) \cite{b}, they seem like glueballs. However,  the
explicit analyses \cite{cam} on them reveal that neither of them
appears to be a pure
meson or a pure glueball. Most probably  they
are mixtures of  glueball and $\bar q q$ meson.
\par The property of the glueball
has been investigated in the lattice gauge theory and in many models based
on the QCD theory. Even in the lattice gauge calculation, there are
different predictions for the $0^{++}$ glueball\cite{w}\cite{ukqcd}\cite{lxq}.
 Some years ago, the mass of
the $0^{++}$ glueball was predicted around $700 - 900$ MeV. Recently, 
IBM group \cite{w} predicts
the lightest $0^{++}$ glueball mass: $(1710\pm63)$  MeV, and UK QCD
group \cite{ukqcd} gives the estimated mass: $(1625\pm92)$ MeV respectively.
The improvement of determination of the $0^{++}$ glueball mass originates
from the more accuracy of the lattice technique, however, at present the
uncertainty still exists .

\par V. A. Novikov {\it et al} \cite{nsvz} first tried to estimate the scalar
glueball mass by using QCD sum rules\cite{svz} , but they only took the 
mass to be
700 MeV by hand because of uncontrolled instanton contributions. Since then,
P. Pascual and R. Tarrach \cite{pt}, S. Narison \cite{n} and J. Bordes {\it et
al} \cite{bgp} presented their calculation on the scalar glueball mass in the
framework of QCD sum rules. They all got a lower mass prediction around
$700 - 900$ MeV when they used the moments $R_{-1}$ or $R_0$ and neglected the
radiative corrections in their calculation of the correlators. 
E. Bagan and T.
Steele \cite{bs} first took account of the radiative corrections in the
correlator calculation. Choosing appropriate moments($R_0$ and $R_1$) 
for their calculation,
they got a higher glueball mass prediction around $1.7$ GeV.  
It seems that the radiative corrections make a big difference on the
prediction of the scalar glueball mass.
Obviously, there are some uncertainties in the determination of the scalar
glueball mass, in order to give the reliable values  
in the QCD sum rules reasonably, an analysis of these uncertainties 
is necessary. 
\par In this paper, we  first give the criteria to choose the
moments, which are obtained by the Borel transformation of the correlator
weighted by different powers of $q^2$, according to application of QCD sum
rules. It is important to choose 
suitable moments to determine the glueball mass\cite{h}. From the criteria
follows that different moment
has different result, but not all of them are reliable.
By choosing appropriate moment, we get the glueball mass without radiative
corrections:
$1.7$ GeV. When the radiative corrections are included in,   
glueball mass shifts a little: $\sim 1.65$ GeV.
\par Secondly , we consider the effect of mixing between
lowest-lying $0^{++}$ glueball and 
$\bar qq$ meson,  i.e. ,
 the gluonic currents and quark currents couple both to
glueball states and ${\bar q}q$ states.  Therefore, there are some exotic
form factors to be determined. By using the low-energy theorem
, we can construct a sum rule for the mixing correlation function
(one gluonic current and one quark current). Through these relationship
and based on the assumption of two states (lowest-lying states
of glueball and ${\bar q}q$ meson) dominance,
we find the mass for $0^{++}$ glueball
is around: $1.9$ GeV, which is a little higher than the
pure resonance prediction while the mass 
for $0^{++}$ meson is around: $1.0$ GeV, which is a little lower than the
pure resonance prediction.
\par The paper is organized as follows. In Sect. 2 a brief review
about the calculation of the mass of physical state from QCD sum rules is given.
In Sect. 3 we discuss the criteria of choosing the moments and the effect of 
the radiative 
corrections. The mixing effect of the glueball with
the meson state is studied in Sect. 4.
Finally, the last section is reserved for a summary.
\section{QCD sum rules and moments}
\indent
\par
Let us consider the correlator
\begin{equation}
\Pi(q^2)={\it i}\int e^{{\it i}qx}\langle0|T\{j(x),j(0)\}|0\rangle dx,        
\end{equation}
where $j(x)$ is the current with definite quantum numbers.
\par  In the deep Euclidean
domain($-q^2\rightarrow\infty$), it is suitable to carry out 
operator product expansion (OPE)
\begin{equation}
\Pi(q^2)=\sum\limits_{n}C_{n}(q^2) O_{n},
\end{equation}
where the $C_{n}(q^2)$ are Wilson coefficients. Then, the correlator 
can be expressed in term of 
 vacuum expectation values of the local operators $O_n$. 
\par On the other hand,  the imaginary part of $\Pi(q^2)$ in
the Minkovski domain(at positive 
values of $q^2$), which is called the spectral density,  is relevant 
with the physical observables. Therefore,  we can extract some information
of the hadrons from QCD calculation by using the 
dispersion relation 
\begin{equation}
\Pi(q^2)=\frac{(q^2)^n}{\pi}
\int\frac{Im\Pi(s)}{s^n(s-q^2)}ds+\sum\limits_{k=0}^{n-1}a_k(q^2)^k ,
\end{equation}
where $a_k$ are some subtraction constants originated from the facial
divergence of $\Pi(q^2)$. In order to  keep control of the convergence 
of the OPE series and enhance the contribution of the lowest lying 
resonance to the spectral density, 
 the standard Borel transformation is used. However, in practice,  
it may be more convenient to  use the moments $R_k$ instead, which 
is defined by 
\begin{eqnarray}\label{moment}
R_k(\tau, s_0)&=&\frac{1}{\tau}\hat{L}[(q^2)^k\{\Pi(Q^2)-\Pi(0)\}]
-\frac{1}{\pi}\int_{s_0}^{+\infty}s^k e^{-s\tau}Im\Pi^{\{pert\}}(s)d
s\\\nonumber
&=&\frac{1}{\pi}\int_{0}^{s_0}s^k e^{-s\tau}Im\Pi(s)d s,  
\end{eqnarray}
where $\hat{L}$ is the Borel transformation and $\tau$ is the Borel 
transformation parameter, $s_0$ is the starting point of
the continuum threshold. Using the higher rank moments, one can 
 enchance the perturbative contribution and suppress resonance
contribution. In the following, we will see the role of $R_k$ in our
analysis.

\section{Criteria of choosing the moments}
\indent
\par In this paper, the $0^{++}$ gluonic current is 
defined as
\begin{equation}\label{current}
j(x)=\alpha_sG_{\mu\nu}^aG_{\mu\nu}^a(x),               
\end{equation}
where $G_{\mu\nu}^a$ in Eq.(\ref{current}) stands for the gluon field strength 
tensor and $\alpha_s$ is the
quark-gluon coupling constant. The current $j(x)$ is the gauge-invariant and
non-renormalization(to two loops order) in pure QCD. 
\par Through operator product expansion,  the
correlator without radiative corrections becomes
\begin{eqnarray}\label{pi}
\Pi(q^2)&=&a_0(Q^2)^2\ln(Q^2/\nu^2)+b_0\langle\alpha_sG^2\rangle\\\nonumber
&+&c_0\frac{\langle g
G^3\rangle}{Q^2}+d_0\frac{\langle\alpha_s^2G^4\rangle}{(Q^2)^2},
\end{eqnarray}
with $Q^2=-q^2>0$, and
\[
\begin{array}{lllllll}
a_0&=&-2(\frac{\alpha_s}{\pi})^2&,&b_0&=&4\alpha_s,\\
c_0&=&8\alpha_s^2&,&d_0&=&8\pi\alpha_s.
\end{array}
\]
For the non-perturbative condensates the following notations and
estimates are used
\begin{eqnarray*}
\langle\alpha_sG^2\rangle&=&\langle\alpha_sG_{\mu\nu}^aG_{\mu\nu}^a\rangle,\\
\langle g G^3\rangle&=&\langle g f_{a b c}G_{\mu\nu}^a G_{\nu\rho}^b
G_{\rho\mu}^c\rangle,\\
\langle\alpha_s^2G^4\rangle&=&14\langle(\alpha_s f_{a b c}
G_{\mu\rho}^a G_{\rho\nu})^2\rangle-\langle(\alpha_s f_{a b c} 
G_{\mu\nu}^a G_{\rho\lambda}^b)^2\rangle.
\end{eqnarray*}
Now,we can apply the standard dispersion representation for the correlator
\begin{equation}
\Pi(Q^2)=\Pi(0)-\Pi^{\prime}(0)+\frac
{(Q^2)^2}{\pi}\int_{0}^{+\infty}\frac{Im\Pi(s)}{s^2(s+Q^2)}d s
\end{equation}
to connect the QCD calculation with the resonance physics. From the low 
energy theorem \cite{nsvz} follows that
\begin{equation}
\Pi(0)=\frac{32\pi}{11}\langle\alpha_sG^2\rangle .
\end{equation}
\par For the physical spectral density $Im\Pi(s)$, one can divide it into
two parts: low energy part and high energy part. Its
high-energy behavior is known as trivial,
\begin{equation}
Im\Pi(s)\longrightarrow\frac{2}{\pi}s^2\alpha_s^2(s),           
\end{equation}
while at low energy region, $Im\Pi(s)$ can be expressed in the single 
narrow width approximation. The single resonance model for $Im\Pi(s)$ leads
\begin{equation}
Im\Pi(s)=\pi f^2M^4\delta(s-M^2),                  
\end{equation}
where M, f are the glueball mass and coupling of the gluon current to the
glueball. Thus we
can proceed the following calculation.
\par
To construct the sum rules,we use the moments $R_k$ defined above, then the 
standard dispersion relation is transformed into
\begin{equation}
R_k(\tau, s_0)=\frac{1}{\pi}\int_{0}^{s_0}s^ke^{-s\tau}Im\Pi(s)ds,
\end{equation}
and from Eq.(\ref{moment}) we have (for $k\geq-1$ )
\begin{equation}
R_k(\tau, s_0)=(-\frac{\partial}{\partial\tau})^{k+1}R_{-1}(\tau,s_0).
\end{equation}
\par Renormalization-group improvement of the sum rules amounts to the 
substitution:
\begin{eqnarray*}
\nu^2 &\rightarrow & \frac{1}{\tau}, \\
\langle g G^3\rangle & \rightarrow & [ \frac{\alpha_s}
{\alpha_s(\nu^2)} ]^{7/11}\langle g G^3\rangle.
\end{eqnarray*}
\par $R_{-1}(\tau,s_0)$ without radiative corrections can be obtained from 
Eq. (\ref{pi}).
\par If we had a complete knowledge of resonances and QCD, we would be
able to fix the glueball mass, then different
moments $R_{k}$ would give the
same result definitely, but we are far from this goal. In practice,
 we cannot calculate the infinite terms in OPE. Therefore, the result will
depend on the choice of the moments. There should be a criteria to
choose some suitable moments at appropriate $s_0$.
As shown in Ref.\cite{bs}, the $R_{-1}$ sum rule leads to a much smaller mass
scale due to the anomalously large contribution of the low-energy part
$\Pi(0)$ of the sum rule and it violates asymptotic freedom at large energy
region. They claimed that $R_{-1}$ was not reliable to predict 
the $0^{++}$ glueball
mass and employed the $R_0$ and $R_1$ moments to predict the $0^{++}$
glueball mass by fitting the stability criteria with the radiative
corrections considered. Their approach showed that the $R_0$ and $R_1$ sum
rules with the radiative corrections result in a higher mass scale
compared to previous mass determination. They didn't analyze how reliable
these moments $R_k$ are for determining the glueball mass. After analyzing
the different moment with the criteria of QCD sum rules, one can find that
$R_0$ is not reliable too for the calculation of $0^{++}$ glueball in the
single narrow width resonance approximation. In order to determine which
moment is the more suitable  and give a reliable mass prediction, we 
re-examine the $R_k$ sum rules.
\par To improve the convergence of the asymptotic series, we study the ratio 
$\frac{R_{k+1}}{R_k}$ , such as
$\frac{R_0}{R_{-1}}$ and $\frac{R_1}{R_0}$ .
In the narrow width approximation, we have
$$M^{2k+4}f^2\exp(-\tau M^2)=R_k(\tau,s_0),$$
and(with $k\geq-1$)
\begin{equation}\label{m}
M^2(\tau,s_0)=\frac{R_{k+1}(\tau,s_0)}{R_k}.
\end{equation}
To proceed calculation, we choose the following parameters
\begin{eqnarray*}
\langle\alpha_sG^2\rangle&=&0.06 GeV^4,\\
\langle gG^3\rangle&=&(0.27 GeV^2)\langle\alpha_sG^2\rangle,\\
\langle\alpha_s^2G^4\rangle&=&\frac{9}{16}\langle\alpha_sG^2\rangle^2,\\
\Lambda_{\bar{MS}}&=&200 MeV,\\
\alpha_s&=&\frac {-4\pi}{11\ln(\tau\Lambda_{\bar{MS}}^2)}.
\end{eqnarray*}
\par $M^2$ and $f^2$ are the functions of $s_0$ which is the starting point of
the continuum threshold, $s_0>M^2$. Since the glueball mass $M$ in
Eq.(\ref{m}) depends on $\tau$ and $s_0$ , we take the stationary point of
$M^2$ versus $\tau$ at an appropriate $s_0$ as the square of the glueball mass.
\par To determine the suitable moment and the appropriate $s_0$, the 
following criteria
are employed: (1), The moments should be chosen to have a balance between
the perturbative and the lowest lying resonance contribution to the sum rule, 
which means
that both the perturbative contribution and the lowest resonance contribution to
the sum rule are dominant in the sum rules; (2),  $s_0$ should be a little 
higher than the physical mass and
approaches it as near as possible due to the continuum threshold hypothesis
and the narrow width approximation; (3), The choice of moments and a suitable 
$s_0$ should
lead to not only a widest flat portions of the plots of $M^2$ versus $\tau$
but also an appropriate parameter region of $\tau$ with the parameter region
compatible to the value of the glueball mass.
 According to these criteria,
the acceptable region of $s_0$ is chosen from  $s_0=3.0$ GeV$^2$ to 
$s_0=4.3$GeV$^2$.

\par 
let's begin our analysis through the $R_k$ sum rules without
radiative corrections. It is known that different
moment has different suppression to the nonperturbative contribution and
the lowest resonance contribution, 
 moments with higher rank enhance the perturbative contribution and 
suppress the lowest resonance contribution to the sum rules.
\par In the sum rule of the moments $R_{-1}$ and $R_0$, although there is
a platform for mass prediction(see Fig. 1), the perturbative
contribution  is less than $30\%$, which is not fit the criteria (1), 
so it is not acceptable.
\par  Using  the moment $R_0$ and $R_1$, one can obtain a balance between
the perturbative 
and the lowest resonance contribution to the sum rules, however 
there is no platform for mass prediction (see Fig. 2). 
It doesn't satisfy the criteria (3), so  this moment is not suitable for
the mass prediction
either. All the previous calculations without radiative corrections were
based on either moment $R_{-1}$ and $R_0$ or moment
$R_0$ and $R_1$, so the results are not very reliable .
\par The ratio $\frac{R_2}{R_1}$ in Fig.
3 gives an excellent platform, and we can find a balance between the
perturbative and the lowest resonance contribution to the sum rules, which 
keep the perturbative contribution 
and the lowest resonance contribution dominant in the sum rules, the moment
$R_1$ satisfies all of the criteria and is reliable for the glueball mass
determination. The curve shows that the $0^{++}$ glueball mass is $1710$ MeV. 
In the acceptable region of $s_0$, the $0^{++}$ glueball mass is
$1710\pm80$GeV. 
\par The moments with higher rank can't stress the lowest resonance 
contribution in the sum rule, because the
higher dimension condensates
will not be negligible(we have little knowledge about
higher dimension
condensates at present). Therefore, we have no way to proceed our
prediction from
$R_k$ with $k>2$.

After taking 
into account radiative corrections, the correlator is\cite{bs}
\begin{eqnarray}
\Pi(q^2)&=&(a_0+a_1\ln(Q^2/\nu^2))(Q^2)^2\ln(Q^2/\nu^2)\\\nonumber
&+&(b_0+b_1\ln(Q^2/\nu^2))\langle\alpha_sG^2\rangle\\\nonumber
&+&(c_0+c_1\ln(Q^2/\nu^2))\frac {\langle gG^3\rangle}{Q^2}+d_0
\frac{\alpha_s^2G^4}{(Q^2)^2}.
\end{eqnarray}                                                                 
where
\begin{eqnarray*}
a_0&=&-2(\frac {\alpha_s}{\pi})^2(1+\frac {51}{4}\frac {\alpha_s}{\pi}),\\
b_0&=&4\alpha_s(1+\frac {49}{12}\frac {\alpha_s}{\pi}),
\end{eqnarray*}
\[
\begin{array}{lll}
c_0=8\alpha_s^2,&d_0=8\pi\alpha_s&,\\
a_1=\frac {11}{2}(\frac {\alpha_s}{\pi})^3,&b_1=-11\frac{\alpha_s^2}{\pi},&
c_1=-58\alpha_s^3.
\end{array}
\]

 The predicted mass
from ratio
$\frac{R_2}{R_1}$ is $\sim 1.65$ Gev(see Fig. 4).
The value is a little lower than the one
without radiative corrections. 
\par In this section, we show how the predicted glueball mass depends on
the choice of the moment. We give the criteria on choosing suitable
moments and
$s_0$ to calculate the glueball mass in QCD sum rules. From the criteria,
only $R_1$ $R_2$ are reliable for  determination of the $0^{++}$ glueball
mass and  the result is
$1.7$ GeV. The radiative corrections do not affect the mass
determination sensitively, they shift the glueball mass a little lower:
$1.65$ GeV. 
\section{Low energy theorem to the mixing picture}
\indent
\par
Now we  proceed to discuss the mixing effect to determination of 
$0^{++}$ glueball mass. Let's consider  the $0^{++}$ quark current with
isospin $I=0$
\begin{equation}
j_2(x)=\frac{1}{\sqrt{2}}(\bar u u(x)+\bar dd(x)) .
\end{equation}
Through operator product expansion, 
 the correlator of the $j_2(x)$ is given by\cite{ryr}
\begin{eqnarray}
\Pi_2(q^2)&=&(a_0'(Q^2)^2\ln(Q^2/\nu^2)+\frac{3}{Q^2}\langle m\bar qq\rangle
+\frac{1}{8\pi
Q^2}\langle\alpha_sG^2\rangle+\frac{b_0'}{(Q^2)^2}\langle\bar qq\rangle^2,
\end{eqnarray}                                                                 
where $Q^2=-q^2>0$, and
\[
\begin{array}{lllllll}
a_0'&=&\frac{3}{8\pi^2}(1+\frac{13\alpha_s}{3\pi})&,&b_0'&=&-\frac{176}{27}\pi
\alpha_s .\\
\end{array}
\]
 The
correlator of the $j_1(x)$ without radiative corrections is not changed.

In order to  estimate the vacuum expectation values of higher dimension  
operators,  the vacuum intermediate states
dominance approximation\cite{svz} has been employed
\begin{eqnarray*}
\langle\bar q\sigma_{\mu\nu}\lambda^aq\bar q\sigma_{\mu\nu}\lambda^aq\rangle&
=&-\frac{16}{3}\langle\bar qq\rangle^2 ,\\
\langle\bar q\gamma_{\mu}\lambda^aq\bar q\gamma_{\mu}\lambda^aq\rangle&
=&-\frac{16}{9}\langle\bar qq\rangle^2 .
\end{eqnarray*}

To proceed the numerical calculation, in addition to the parameters we have 
chosen above, the following parameters are taken
\begin{eqnarray*}
\langle\bar qq\rangle&=&-(0.25GeV)^3,\\
\langle m\bar qq\rangle&=&-(0.1GeV)^4,\\
\alpha_s&=&0.28,
\end{eqnarray*}
where the scale of the running coupling is set at the glueball mass.
\par Through the $R_k$ defined above, we can get the corresponding moments
$R_k$ and $R_k'$ for $\Pi(q^2)$ and $\Pi_2(q^2)$
\begin{eqnarray}
R_{0}(\tau,s_0)&=&-\frac{2a_0}{\tau^3}\lfloor1-\rho_2(s_0\tau)\rfloor
+c_0\langle gG^3\rangle+d_0\langle \alpha_s^2G^4\rangle\tau,\\
R_{1}(\tau,s_0)&=&-\frac{6a_0}{\tau^4}\lfloor1-\rho_3(s_0\tau)\rfloor-d_0
\langle\alpha_s^2G^4\rangle,\\
R_{2}(\tau,s_0)&=&-\frac{24a_0}{\tau^5}\lfloor1-\rho_4(s_0\tau)\rfloor,\\
R_{0}'(\tau,s_0)&=&\frac{a_0'}{\tau^2}\lfloor1-\rho_1(s_0\tau)\rfloor+3\langle
m\bar qq\rangle
+\frac{1}{8\pi}\langle\alpha_sG^2\rangle+b_0'\tau\langle\bar qq\rangle^2 ,\\
R_{1}'(\tau,s_0)&=&\frac{2a_0'}{\tau^3}\lfloor1-\rho_2(s_0\tau)\rfloor-b_0'
\langle\bar
qq\rangle^2,
\end{eqnarray}
where
\begin{equation}
\rho_k(x)\equiv e^{-x}\sum\limits_{j=0}^{k}\frac{x^j}{j!}.
\end{equation}

By using the Low-energy theorem \cite{al}, we can construct another correlator 
for the quark current with the gluonic current 
\begin{equation}\label{mix}
\lim_{q\rightarrow 0}{\it i}\int dxe^{{\it
i}qx}\langle 0|
T[ \frac{1}{\sqrt{2}}(\bar uu+\bar dd),\alpha_s G^2] |0\rangle=
\frac{72\sqrt{2}\pi}{29}\langle \bar uu\rangle,
\end{equation}

In order to factorize the spectral density, we define the couplings of the
currents to the physical states in the following way
\begin{eqnarray}
\langle 0|j_1|Q\rangle=f_{12}m_2&,&\langle
0|j_1|G\rangle=f_{11}m_1,\\\nonumber
\langle 0|j_2|Q\rangle=f_{22}m_2&,&\langle 0|j_2|G\rangle=f_{21}m_1,
\end{eqnarray}
where $m_1$ and $m_2$ refer to the glueball(including few part of quark
component) mass and the $\bar qq$ meson(including few part of
gluon component) mass, $|Q\rangle$ 
 and $|G\rangle$ 
  refer
to the $\bar qq$ meson state and  the glueball state respectively.
\par We indicate that the gluon current couples to both the glueball and
quark states, so does the quark current. In the real physical world, the
physical state is not pure glueball state or quark state, the mixing
effect should not be omitted without any reanonable argument.
After choosing the two resonances plus continuum state approximation, the 
spectral 
density of the currents of $j_1(x)$ and $j_2(x)$ read in following  
respectively
\begin{eqnarray}
Im\Pi_1(s)&=&m^2_2f^2_{12}\delta(s-m^2_2)+m^2_1f^2_{11}\delta(s-m^2_1)+
\frac{2}{\pi}s^2\alpha^2_s\theta(s-s_0),\\
Im\Pi_2(s)&=&m^2_2f^2_{22}\delta(s-m^2_2)+m^2_1f^2_{21}\delta(s-m^2_1)+a_0's
\theta(s-s_0) .
\end{eqnarray}

Then it is straightforward to get the moments
\begin{eqnarray}
R_0&=&\frac{1}{\pi}\{m^2_2e^{-m^2_2\tau} f^2_{12}+m^2_1e^{-m^2_1\tau} f^2_{11}
\},\\
R_1&=&\frac{1}{\pi}\{m^4_2e^{-m^2_2\tau} f^2_{12}+m^4_1e^{-m^2_1\tau} f^2_{11}\}
,\\
R_2&=&\frac{1}{\pi}\{m^6_2e^{-m^2_2\tau} f^2_{12}+m^6_1e^{-m^2_1\tau}
f^2_{11}\},\\
R_0'&=&\frac{1}{\pi}\{m^2_2e^{-m^2_2\tau} f^2_{22}+m^2_1e^{-m^2_1\tau} f^2_{21}\},\\
R_1'&=&\frac{1}{\pi}\{m^4_2e^{-m^2_2\tau} f^2_{22}+m^4_1e^{-m^2_1\tau} f^2_{21}\}.
\end{eqnarray}

In the meanwhile, assuming the states $|G\rangle$ and $|Q\rangle$ saturate
the l.h.s of Eq. (\ref{mix}), we can obtain 
\begin{equation}
\lim_{q\rightarrow 0}{\it i}\int dxe^{{\it
i}qx}\langle 0|
T[ \frac{1}{\sqrt{2}}(\bar uu+\bar dd),\alpha_s G^2] |0\rangle=
f_{22}f_{12}+f_{21}f_{11}.
\end{equation}

The next step is to equate the QCD side with the
hadron side one by one, and we get
a set of equations. 
Giving various of reasonable parameters $s_0$ and 
$\tau$ and through solving this series of equations,  we can get
a series of  the two states' masses. We illustrate our result in Fig.5.
In this figure, the solid line
corresponding to the glueball and the doted line corresponding to the meson,
the points of the plateau compatible to the parameters are regarded as the
mass prediction points. We find
that $s_0=3.7$ GeV$^2$ is the best favorable value for $s_0$. There is no
platform for $\tau$ above $0.6$ GeV$^{-2}$, we can read the masses 
prediction: glueball with mass around $1.9$ GeV 
and meson with mass around $1.0$ GeV.
We find that the glueball mass a little higher than the pure glueball
state while the quark state mass is a little lower than the pure quark
state.
\section{Summary}
\indent
\par In this paper, we analyze the determination of the scalar glueball 
mass based on the duality among resonance physics and QCD. 
The modified Borel transformation has been
employed, it makes the calculation more convenient and reasonable.
\par   We first conclude that it is important to choose suitable moments
for the determination of $0^{++}$ glueball mass.  To stress
the contribution of the lowest resonance and make the
perturbative contribution dominant in sum rules, the 
criteria on the choice of the moment and  continuum threshold are given.
These criteria make
it reliable to choose a suitable moment for the calculation of the glueball
mass. We find moments $R_{-1}$, $R_0$ and
$R_k$ with higher rank $k>2$ aren't suitable for the mass determination in
the single narrow width resonance approximation.
 The ratio of moments $\frac{R_2}{R_1}$ is the most preferable for
the determination of $0^{++}$
glueball mass. The numerical calculation shows that the mass is around
$1.7$ GeV without radiative corrections.
\par  When the radiative correction is take into account, it shifts
to $1.65$ GeV. 

\par Secondly, we consider the physical states as composite resonances,
which include both gluon component and quark component, so
we saturate the spectral density with
two physical resonances, in this way we consider not only the couplings 
of gluonic current to both
glueball state and quark state, but also the couplings of quark current to
 quark state and glueball state. Employing the Low-energy theorem and
different moments, we predict the masses of  glueball and normal meson 
 from a set
of coupled equations: glueball mass is around $1.9$ GeV, which 
is a little higher than the one without mixing($\sim 1.7Gev$), 
while  mass of the quark state  is around $1.0 GeV$. 
 a little lower than the pure quark state($\sim 1.1Gev$). We 
conclude that the mixing between the glueball and the quark state is not
large.
\\

{\bf Acknowledgment}

This work is supported in part by the national natural science foundation
of P. R. China.

\newpage
\par
{\large\bf Figure caption}
\par
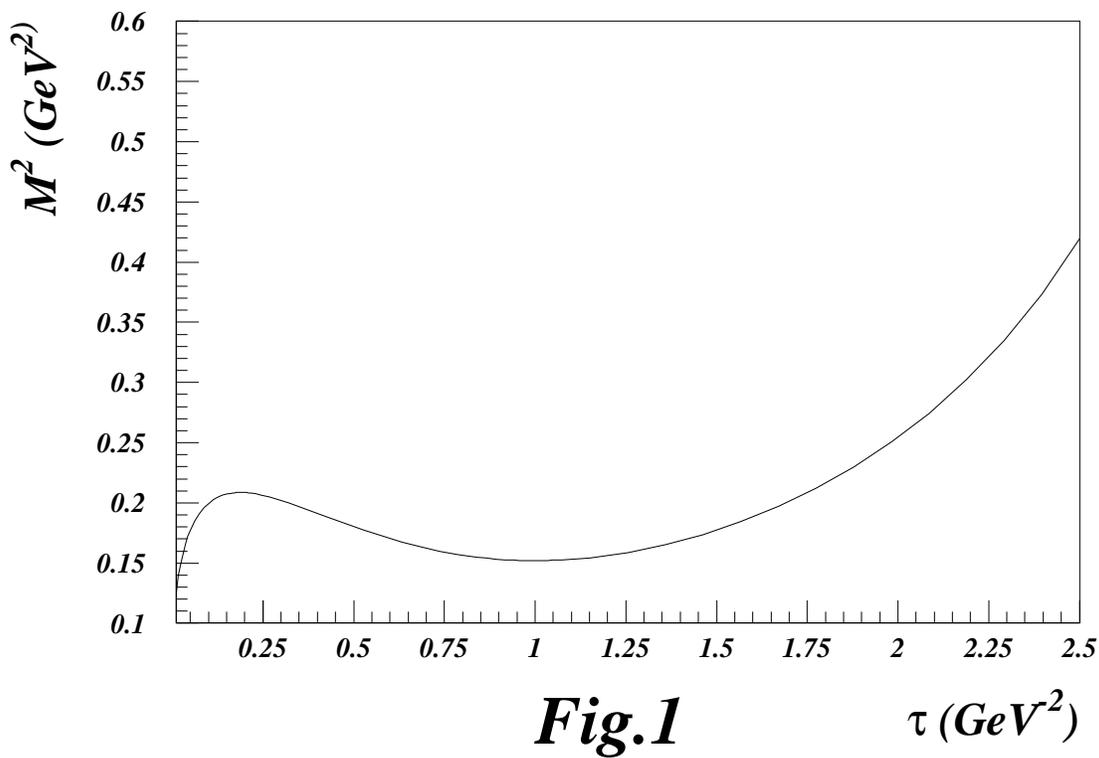
Figure 1: $\frac{R_0}{R_{-1}}$ versus $\tau$ at $s_0=3.6$ GeV$^2$ without
radiative corrections.
\par
\par
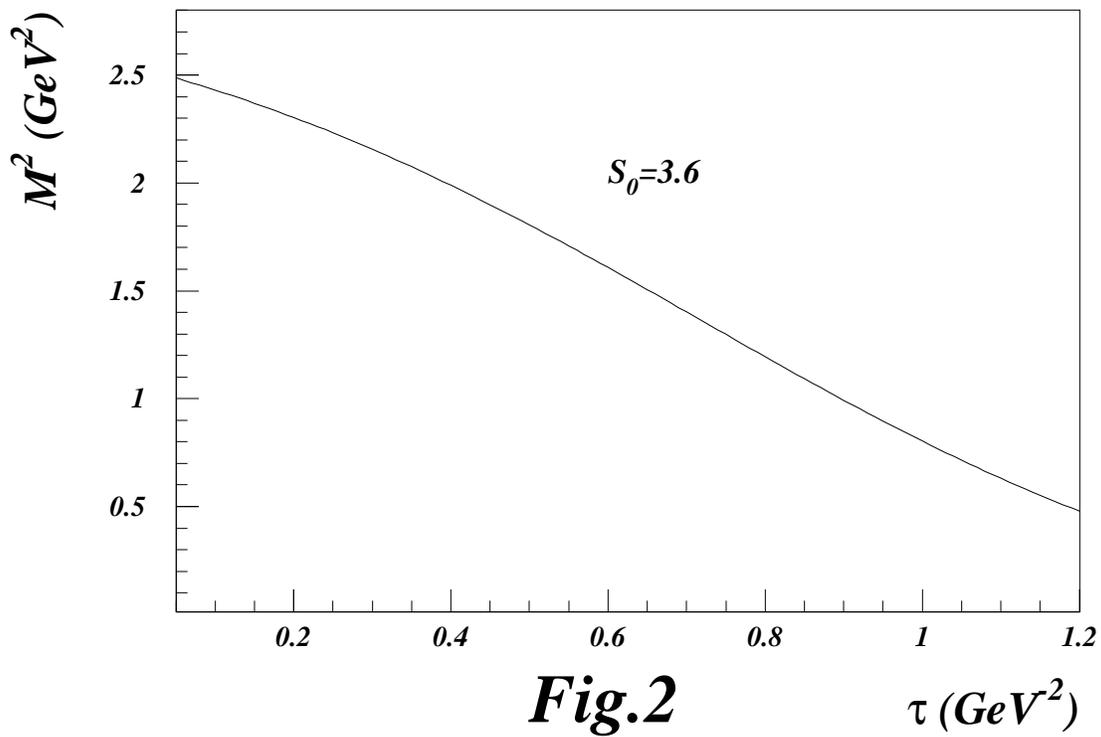
Figure 2: $\frac{R_1}{R_0}$ versus $\tau$ at $s_0=3.6$ GeV$^2$ without
radiative corrections.
\par
\par
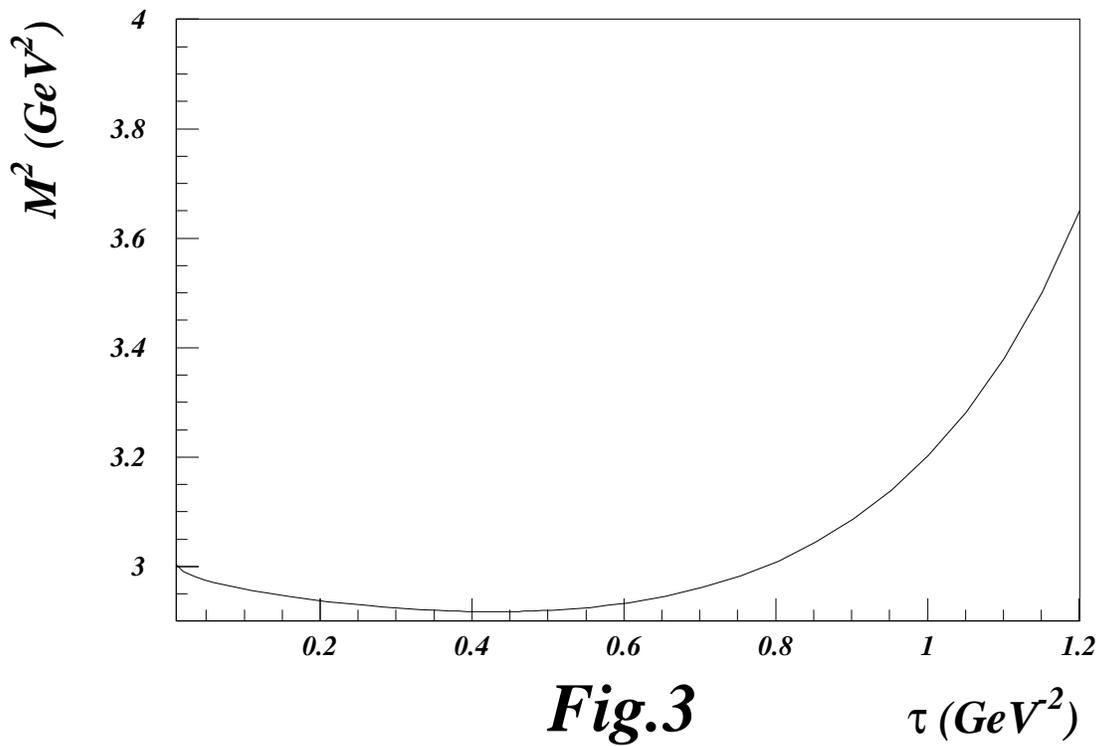
Figure 3: $\frac{R_2}{R_1}$ versus $\tau$ at $s_0=3.6$ GeV$^2$ without
radiative corrections.
\par 
\par
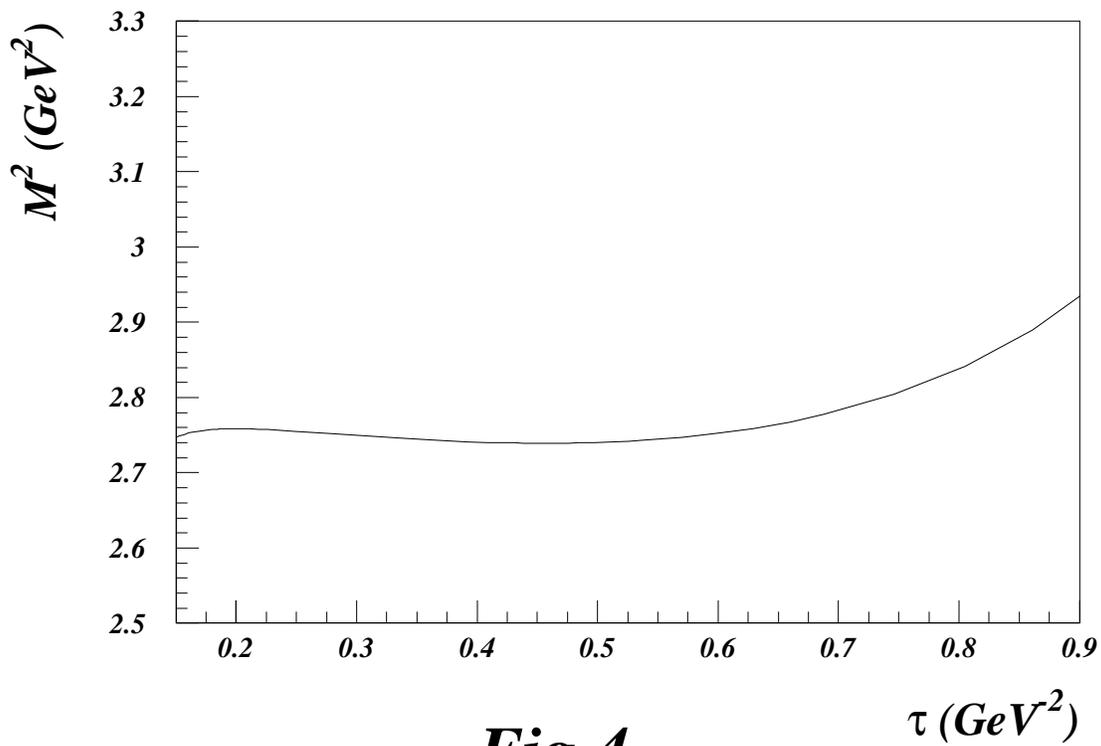
Figure 4: $\frac{R_2}{R_1}$ versus $\tau$ at $s_0=3.6$ GeV$^2$ with radiative
corrections.
\par
\par
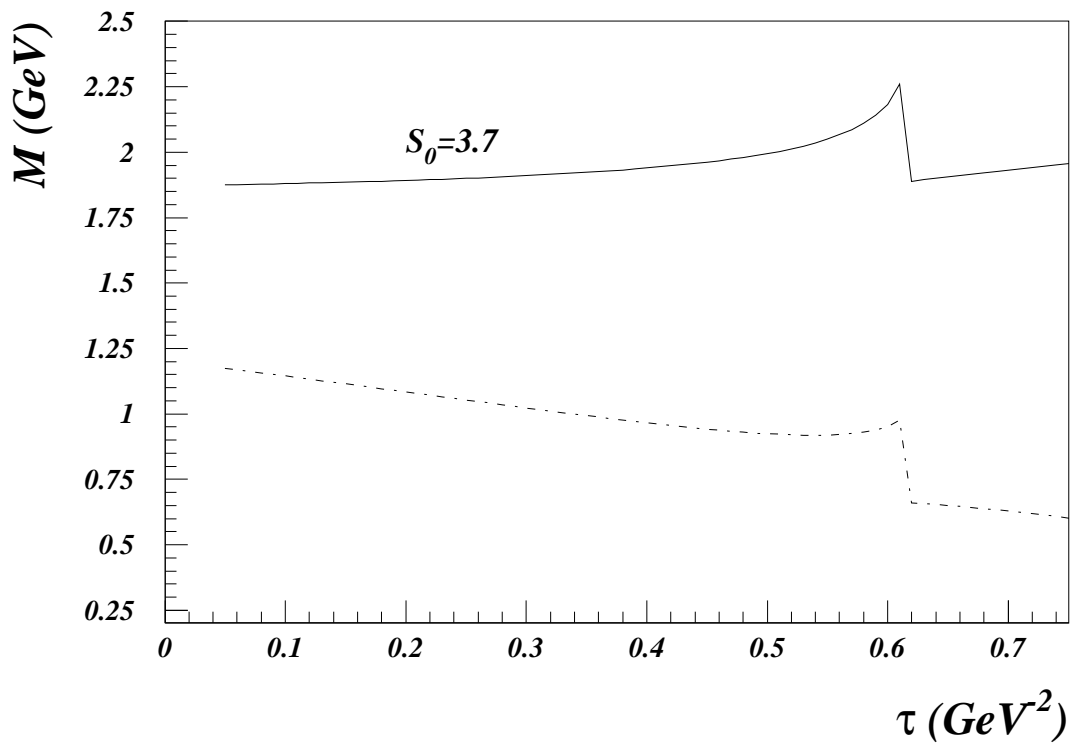
Figure 5: $M$ versus $\tau$ at $s_0=3.7$ GeV$^2$.


\begin{thebibliography}{15}
\bibitem{cbc}
S. Spanier, hep-ex/9801006, (1998).
\bibitem{b}
D. V. Bugg, {\it et al}. Phys. Lett. B{\bf 353}, 378 (1995).
\bibitem{cam}
Curtis A. Meyer, hep-ex/9707008, (1997).
\bibitem{w} 
D. Weingarten, Nucl. Phys. B (Proc. Suppl.){\bf 34}, 29 (1994).
\bibitem{ukqcd}
G.Bali {\it et al}. (UKQCD), Phys. Lett. B{\bf 309}, 378 (1993).
\bibitem{lxq}
Xiang-Qian Luo, {\it et al}, Nucl. Phys. Proc. Suppl. {\bf 53}, 243 (1997).
\bibitem{nsvz}
V. A. Novikov, M. A. Shifman, A. I. Vainshtein and V. I. Zakharov, Nucl. Phys.
B{\bf 165}, 67 (1980).
\bibitem{svz}
M. A. Shifman, A. I. Vainshtein and V. I. Zakharov, Nucl. Phys.
B{\bf 147}, 385 (1979).
\bibitem{pt}
P. Pascual and R. Tarrach, Phys. Lett. B{\bf 113}, 495 (1982).
\bibitem{n}
S. Narison, Z. Phys. C{\bf 26}, 209 (1984).
\bibitem{bgp}
J. Bordes, V. Gim\`{e}nez and J. A. Pe\~{n}arrocha, Phys. Lett.
B{\bf 223}, 251 (1989).
\bibitem{bs}
E. Bagan and T. G. Steele, Phys. Lett. B{\bf 243}, 413 (1990).
\bibitem{h}
S. Narison, Nucl. Phys. B{\bf 509}, 312(1998).\\
Tao Huang, Ailin Zhang, hep-ph/9801214.
\bibitem{ryr}
L. J. Reinders, S. Yazaki and H. R. Rubinstein , Nucl. Phys.
B{\bf 196}, 125 (1982).
\bibitem{al}
V. A. Novikov, M. A. Shifman, A. I. Vainshtein and V. I. Zakharov, Nucl. Phys.
B{\bf 191}, 301 (1981).

\end{thebibliography}
\end{document}